\documentclass{article}
\usepackage[utf8]{inputenc}
\usepackage{amsmath}
\usepackage{amsfonts}
\usepackage{bm}
\usepackage{verbatim}
\usepackage{MnSymbol}
\usepackage{graphicx}
\usepackage{subfig}
\usepackage{lineno,hyperref}
\usepackage[utf8]{inputenc}
\usepackage[T1]{fontenc}
\usepackage{nomencl}
\usepackage{bm}
\usepackage{verbatim}
\usepackage{MnSymbol}
\makenomenclature
\usepackage{etoolbox}
\usepackage{amsfonts}
\usepackage{bbold}
\usepackage{algorithm}
\usepackage{algcompatible}
\usepackage{authblk}
\usepackage{algorithm}
\usepackage{multirow}
\usepackage{enumerate}
\usepackage{xcolor}
\usepackage{soul}
\usepackage{empheq} 

\newcommand{\argmax}{\arg\!\max}

\title{Scalable Multi-agent Reinforcement Learning for Factory-wide Dynamic Scheduling}
\author[1]{Jaeyeon Jang}
\author[2]{Diego Klabjan}
\author[3]{Han Liu}
\author[4]{Nital S. Patel}
\author[4]{Xiuqi Li}
\author[4]{Balakrishnan Ananthanarayanan}
\author[4]{Husam Dauod}
\author[2]{Tzung-Han Juang}

\affil[1]{Department of Data Science, The Catholic University of Korea}
\affil[2]{Department of Industrial Engineering and Management
Sciences, Northwestern University}
\affil[3]{Department of Computer Science, Northwestern University}
\affil[4]{Intel Corporation}

\begin{document}

\maketitle

\begin{abstract}
\noindent Real-time dynamic scheduling is a crucial but notoriously challenging task in modern manufacturing processes due to its high decision complexity.  Recently, reinforcement learning (RL) has been gaining attention as an impactful technique to handle this challenge. However, classical RL methods typically rely on human-made dispatching rules, which are not suitable for large-scale factory-wide scheduling. To bridge this gap, this paper applies a leader-follower multi-agent RL (MARL) concept to obtain desired coordination after decomposing the scheduling problem into a set of sub-problems that are handled by each individual agent for scalability. We further strengthen the procedure by proposing a rule-based conversion algorithm to prevent catastrophic loss of production capacity due to an agent's error. Our experimental results demonstrate that the proposed model outperforms the state-of-the-art deep RL-based scheduling models in various aspects. Additionally, the proposed model provides the most robust scheduling performance to demand changes. Overall, the proposed MARL-based scheduling model presents a promising solution to the real-time scheduling problem, with potential applications in various manufacturing industries.
\end{abstract}

\section{Introduction}
Scheduling in semiconductor manufacturing is an exceptionally challenging task due to several complex characteristics. These include fluctuating demand, a variety of product types, intricate precedence constraints of operations, hundreds of machines spread across numerous workstations, a high frequency of re-entrant phenomena, frequent machine breakdowns and maintenance, and a vast number of possible routes for a job \cite{Shiue2021, Kim2017a, Yu2021, Chen2019a}. This complicated and challenging scheduling task has been represented, at a high level of abstraction, as a flexible job shop scheduling problem (FJSSP)\cite{Gao2015, Luo2020} and diverse methods have been proposed to tackle the FJSSP \cite{Fattahi2007, Amjad2018, Gao2019}.

With the increasing complexity and uncertainty of modern manufacturing systems, the FJSSP has garnered considerable attention in recent years \cite{Gao2019}. This problem builds upon the conventional job shop scheduling problem by increasing flexibility and universality to address advanced manufacturing systems such as automobile assembly, chemical material processing, and semiconductor manufacturing \cite{Guo2016, Gao2019a}. In the FJSSP, each operation can be processed on multiple compatible machines, each with a different processing time. As a result, the FJSSP is a formidable NP-hard problem \cite{Kacem2002}.

Many algorithms have been developed to address the FJSSP problem \cite{Gao2015, Fattahi2007, Amjad2018, Gao2019}. However, most of them rely on an unrealistic assumption of a static scheduling environment where all the relevant information about the production environment is known in advance, and all future events are controlled as planned. The aim of these algorithms is to create the best deterministic scheduling plan without considering any future modifications. However, in complex manufacturing systems, dynamic events such as machine breakdowns, unscheduled maintenance, job insertion, and modification of order have become increasingly common \cite{Ouelhadj2009}. As a result, a deterministic schedule is not usually executed as planned, deteriorating the production efficiency significantly. In addition, factory scheduling involves many hard operational constraints due to the complexity and uncertainty of semiconductor manufacturing systems. Therefore, developing a real-time scheduling method capable of handling these constraints is of utmost importance. Recently, reinforcement learning (RL) has achieved huge attention in this area due to its ability to learn a good policy by trial and error in a complex environment, thereby enabling real-time decision-making \cite{Luo2020, Liu2020, Chang2022, Min2022}.

Dynamic FJSSP (DFJSSP) has been extensively studied over the past decade, with dispatching rules and metaheuristics being commonly used techniques \cite{Lou2012}. Dispatching rules are popular among practitioners due to their simplicity and time efficiency. However, their solutions may not be of high quality since no single best rule can be applied to all situations \cite{Min2022}. In contrast, metaheuristics can provide higher quality solutions by breaking down the dynamic scheduling problem into a series of static sub-problems \cite{Luo2020}. Nevertheless, metaheuristics algorithms, such as genetic algorithm, tabu search, and particle swarm optimization, may suffer from significant computational cost due to their huge search spaces \cite{Nilsen2016, Ning2016}. Thus, their practicality is questionable.

With the rapid advancement of artificial intelligence, deep RL (DRL) has emerged as an important approach in real-time scheduling due to its practicality in adapting to numerous dynamic events in real-time \cite{Cunha2020}. Despite its performance improvement over dispatching rules and metaheuristics, existing DRL-based methods have several limitations. Firstly, most of them heavily rely on dispatching rules. Secondly, prior work does not consider hard operational constraints in dynamic environments which is crucial for a factory-wide production environment. 

The factory-wide DFJSSP presents a challenge for obtaining optimal or near-optimal solutions due to its complex nature and numerous hard constraints. Even the introduction of DRL cannot address this challenge as the entire search space becomes impractical \cite{Subramanian2016}. To overcome this limitation, we introduce the concept of a multi-agent RL (MARL) approach \cite{Zhang2021a} that decomposes the problem into a set of operation-specific sub-problems. In this concept, achieving team goals requires effective coordination among all agents \cite{Foerster2018, Rashid2018, Yang2020}. However, in the context of a factory-wide production environment, guiding multiple agents with commonly used shared team reward \cite{Tan1993} is not effective due to the complex hierarchy and intricate interconnections of numerous sequential operations. Therefore, instead of a shared reward, we use operation-wise rewards that allow each agent to focus on its assigned operation, thus significantly reducing the search space. However, cooperation may not be achieved through this approach, even if each agent performs well in its assigned operation. To address this challenge, we introduce a new outer agent proposed in \cite{jang2023learning}, called the leader, which coordinates the agents assigned to each operation, which we call followers, by providing an abstract goal for communication.

In a real factory production environment, converting from one product type to another requires a significant operational cost. Therefore, strict constraints are imposed to ensure that conversions are carefully executed. Since there is limited flexibility in the conversion process, even one bad decision made by a follower can have catastrophic consequences. For instance, machines can unexpectedly become idle, resulting in a significant loss of production capacity over a long period of time. However, it is impossible to guarantee that RL agents will not make any errors. Therefore, in situations where a follower's bad decision can significantly impact performance, we implement a rule-based conversion algorithm. This allows us to apply a set of predetermined rules that supersede the follower's decision-making process. In this way, we can prevent potential performance deterioration that may occur due to a follower's decision.

We benchmarked the proposed model by comparing it with the state-of-the-art DRL-based scheduling models utilizing dispatching rules in real factory environments. Experimental results show that existing DRL-based models cannot effectively learn a scheduling strategy in complex real factory environments. This reveals that a combination of predefined dispatching rules is insufficient. In contrast, our proposed model achieved significant performance improvement during training, resulting in much better scheduling strategies. Furthermore, our proposed model was found to be more robust to increasing demand, as its performance deterioration was relatively small. Specifically, compared to the existing models in the most challenging situation with high demand, the proposed model reduces tardiness by 10.4\% and improves the completion rate by 31.4\% on average.  Our main contributions are as follows.
\begin{itemize}
\item An RL-based scheduling model that is not based on human-made dispatching rules is proposed.
\item An MARL concept based on leader-follower that utilizes abstract goals for cooperation among followers is successfully introduced to obtain scalability for large-scale factory-wide dynamic scheduling problems.
\item A rule-based conversion algorithm is developed and incorporated to prevent catastrophic loss of production capacity.
\end{itemize}

While the methodology is based on \cite{jang2023learning} herein we modify it by incorporating features specific to real-world factory environments, including scheduled maintenance and unexpected machine breakdowns. Additionally, we omit the concept of synthetic rewards present in \cite{jang2023learning} and instead developed an operation-wise rewarding strategy alongside a rule-based conversion algorithm to prevent catastrophic loss of production capacity. The remainder of this paper is organized as follows. In Section \ref{sec:related}, related works on RL-based dynamic scheduling are overviewed. Section \ref{sec: PF} formulates factory-wide DFJSSP, the target scheduling problem of this paper. In Section \ref{sec:model}, we detail the proposed model, incorporating the leader-follower MARL algorithm from \cite{jang2023learning}, which we integrate into our scheduling model. This approach is utilized to coordinate multiple agents in factory-wide scheduling, a problem characterized by a directed acyclic graph (DAG). In Section \ref{sec:experiment}, the proposed model is thoroughly benchmarked with various baselines including several state-of-the-art models. Finally, Section \ref{sec:conclusion} concludes this study by discussing some limitations and future research directions.

\section{Related works} \label{sec:related}
RL has proven to be successful in various fields, including robot control \cite{Cui2017}, manufacturing \cite{Hein2018}, and gaming \cite{Rashid2018} in the past few decades. As a result, many researchers in the scheduling field have become interested in this learning technique \cite{Luo2020, Liu2020, Chang2022, Min2022}.  They have adopted the concept of training adaptable agents to choose the best action (usually dispatching rules) at decision points through numerous trial and error. More specifically, dynamic scheduling problems have traditionally been approached as Markov decision processes (MDPs), which are solved using classical RL algorithms such as Q-learning and SARSA \cite{RichardS.Sutton2018}. For example, Aydin and {\"{O}}ztemel \cite{Aydin2000} applied the Q-learning algorithm to train an agent that selects the most appropriate rule among three popular dispatching rules: the shortest processing time (SPT) rule, the cost over time rule, and the critical ratio rule. Their goal was to determine the next job in real-time, with the aim of minimizing mean tardiness. Similarly, Wei and Mingyang \cite{Wei2005} introduced an agent trained using Q-learning to select the best rule at each decision point after designing several composite dispatching rules. In \cite{Wang2005}, Q-learning was used to select the best dispatching rule among the FIFO rule, the earliest due date rule, and the SPT rule based on the number of waiting jobs and the estimated total lateness. Chen et al. \cite{Chen2010} used Q-learning to produce a single composite rule by subsequently combining several dispatching rules based on a weighted aggregation method. Finally, Shahrabi et al. \cite{Shahrabi2017} utilized Q-learning to obtain optimal parameters for the variable neighborhood search algorithm, a heuristic method that searches for the best schedules in dynamic job shop problems with unequal job arrival times and machine failures. While conventional RL-based scheduling methods have made significant progress, the latest production systems have presented more scheduling challenges, requiring exploration of larger state spaces \cite{Luo2021} and making the conventional RL approaches less effective.

In advanced production systems, there are almost infinite numbers of different states, making it difficult for conventional RL algorithms to obtain a good solution within a limited time budget, which raises questions about their practicality. However, with the emergence of deep learning, DRL has made breakthroughs in real-time scheduling with large and continuous state spaces by using deep neural networks (DNNs) to estimate the value function and/or policy \cite{Mnih2015, Lillicrap2016}. Consequently, recent developments in the field of scheduling have largely been achieved through DRL. For instance, Waschneck et al. \cite{Waschneck2018} applied the concept of MARL while using a deep Q-network (DQN) so that agents can determine the best dispatching rules for a work center to maximize machine utilization. Hu et al. \cite{Hu2020} proposed a graph convolutional network-based DQN to address route flexibility and stochastic arrivals of products in flexible manufacturing systems. Liu, Chang, and Tseng \cite{Liu2020} introduced an actor-critic architecture while applying a parallel training method that combines asynchronous updates and deep deterministic policy gradient (DDPG). In this work, multiple agents were optimized to determine the best rule among several simple dispatching rules for each machine. Luo \cite{Luo2020} designed six composite dispatching rules and proposed double DQN (DDQN) to determine the best rule among the designed dispatching rules at every rescheduling point to minimize total tardiness. In the following work, Luo et al. \cite{Luo2021} extended Luo's model to a two-hierarchy deep Q-network by introducing a higher-level DDQN that determines the optimization goal. Chang et al. \cite{Chang2022} enhanced DDQN by proposing a soft $\epsilon$-greedy behavior policy to balance exploration and exploitation but still require the trained agent to perform the same task of choosing a dispatching rule from a set of rules. Min and Kim \cite{Min2022} used a deep deterministic policy gradient algorithm to dynamically tune the parameters of the apparent tardiness cost dispatching rule, which is known to yield schedules that minimize the total weighted tardiness \cite{Vepsalainen1987}. 

While the existing methods have been effective for solving dynamic scheduling problems, there are weak points. Specifically, they assume that there are no restrictions on job selection and that numerous product type conversions can be executed. However, in reality, conversion is significantly limited to minimize operational costs and maximize machine utilization, while also considering operational constraints within a factory. As a result, these existing scheduling methods cannot be applied to real-world factory environments or may significantly impair scheduling performance. Thus, this study proposes a MARL based scheduling algorithm that specifically targets real-world factory environments.

\section{Problem statement} \label{sec: PF}
The factory-wide DFJSSP assumes that jobs with the same product type have identical unit processing times for an operation. We also assume earlier arriving jobs are processed first when multiple jobs with the same product type are waiting to be processed. When a machine undergoes a change in operation mode or product type, a conversion time depending on the previous operation/product type and the new one is needed.

We describe factory-wide DFJSSP as a planning problem consisting of a period of $N$ shift. Given a product type $p$, we denote $K_p$ to be the number of lots of demand and $J_p$ to be the number of required operations. To produce a product $p$, we need to sequentially conduct a sequence of operations $\{O_{p,j}|j=1, 2, \cdots, J_p\}$ where $O_{p,j}$ is the $j$-th operation of $p$. We can perform each operation $O_{p,j}$ on a set of compatible machines $\bm{M}_{p,j} \subseteq \bm{M}$, where $\bm{M}$ is the set of all machines. In general, most machines have the ability to produce multiple type of products and perform multiple operations. However, changing operations or product types requires conversion time. To this end, we denote $CO_{p_0,o_0,p_1,o_1}$ to be the time required to convert from operation $o_0$ for product type $p_0$ to operation $o_1$ for product type $p_1$. To better reflect a real-world factory environment, we also impose a conversion time threshold $TH$ on all machines for each shift. This is necessary since excessive conversion time can have negative impacts on machine utilization and productivity. Specifically, in a given shift of length $S$, the total conversion time for each machine cannot exceed $TH$. The time required to process the $k$-th lot of product type $p$ on operation $O_{p,j}$ is the product of the number of units in the lot $U_{p,k}$ and the unit processing time $PR_{p,j}$.

In real-world factory environments, both scheduled maintenance to prevent machine breakdowns and unscheduled maintenance due to unexpected breakdowns must be considered. For machine $l$, $SM_{l,t}$ is defined as 1 if $l$ is under scheduled maintenance at time $t$, and 0 otherwise. Additionally, $\bm{\Omega}_t$ represents the set of machines undergoing maintenance due to unexpected breakdowns, which is really a random event. For simplicity, we assume that unscheduled maintenance begins only after the current lot is processed.

Given the provided setting, we list the decision variables in factory-wide DFJSSP.
\begin{itemize}   
\item $X_{p,j,k,l,t}=\begin{cases}
1, &\text{if $O_{p,j}$ is assigned on $l \in \bm{M}_{p,j}$ for the $k$-th lot of product $p$ at $t$} \\
0, & \text{otherwise}\\
\end{cases}$
\item  $C_{p,j,k}$: Completion time of $O_{p,j}$ for the $k$-th lot of product type $p$
\item  $PT_{l,t}$: product type setup of $l \in \bm{M}$ at $t$
\item  $OP_{l,t}$: operation setup of $l \in \bm{M}$ at $t$
\item $Q_{l,t}=\begin{cases}
1, &\text{if $l \in \bm{M}$ is not idle  at $t$}\\
0, & \text{otherwise}\\
\end{cases}$
\end{itemize}

\noindent Variable $X_{p,j,k,l,t}$ determines which machine performs $O_{p,j}$ for the $k$-th lot of product type $p$, while $Q_{l,t}$ characterizes the status of $M_l$. Specifically, $Q_{l,t}$ is set to 1 if $l$ is currently processing, undergoing a setup change, or undergoing scheduled or unscheduled maintenance; otherwise, it is set to 0, as detailed in (\ref{eq2}). Whenever there is an idle machine, a decision should be promptly made for that machine. For example, if $Q_{l,t}=0$ for an $l \in \bm{M}_{p,j}$, and $(k-1)$-th lot of $p$ has already been processed in $O_{p,j}$, we can set $X_{p,j,k,l,t}=1$ to initiate $k$-th lot of $p$ using $l$ for $O_{p,j}$. Conversely, if no machines are available at time $t$, the decision is set by fixing $X_{p,j,k,l,t}=0$. Formally,
\begin{equation}\label{eq0}
X_{p,j,k,l,t} \leq (1-Q_{l,t})\bm{1}_{\bm{M}_{p,j}}(l)\sum_{l \in \bm{M}_{p,j}} \sum_{\tau=1}^{t-1}X_{p,j,k-1,l,\tau}.
\end{equation}

Additionally, let $D_{p,k}$ be the due time of lot $k$ of product $p$. Our goal is to identify decision variables that minimize the number of delayed jobs, or equivalently, maximize the completion rate, within the planning horizon of $N$ shifts, where each shift comprises of $S$ decision points. In this paper, for simplicity, we assume that lots within the same product type contain nearly identical unit counts. Consequently, our objective function, formalized as (\ref{eq1}), does not account for variations in unit numbers among lots of the same product type. In the following, the expectation is with respect to unscheduled maintenance captured by random events $\Omega_t$. Equations and constrains (\ref{eq0}) and (\ref{eq2}) - (\ref{eq11}) hold for any realization of $\Omega_t$.

\begin{equation}\label{eq1}
\text{Minimize}\quad \mathbb{E}\left[\sum_{p}\sum_{k=1}^{K_p}H_{p,J_p, k}\right],\:\text{where}\: H_{p,J_p, k}=\begin{cases}
1, &\text{if $D_{p,k}<C_{p,J_p,k}$}\\
0, & \text{otherwise}\\
\end{cases}\\
\end{equation}

\begin{align}\label{eq2}
\text{subject to}\nonumber\\
Q_{l,t}=&\bm{1}_{\bm{\Omega}_t}(l)SM_{l,t}\sum_{p}\sum_{k=1}^{K_p}\sum_{j=1}^{J_p}\bm{1}_{\bm{M}_{p,j}}(l)(\sum_{\tau=t-PR_{p,j}U_{p,k}}^{t}{Z_{p,j,l,\tau}X_{p,j,k,l,\tau}}\nonumber\\+&\sum_{\tau=t-(P_{p,j}U_{p,k}+{CO}_{{PT}_{l,t},{OP}_{l,t},p,O_{p,j}})}^t{(1-Z_{p,j,l,\tau})X_{p,j,k,l,\tau}})\forall l,t,
\end{align}

\begin{align*}
\text{where}\: Z_{p,j,l,t}=\begin{cases}
1, &\text{if $PT_{l,t}=p$ and $OP_{l,t}=O_{p,j}$}\\
0, & \text{otherwise}\\
\end{cases},
\end{align*}

\begin{equation}\label{eq3}
Q_{l,t}\le 1 \forall l,t,
\end{equation}

\begin{equation}\label{eq4}
PT_{l,t}=\begin{cases}
p, &\text{if $X_{p,j,k,l,t}=1$}\\
PT_{l,t-1}, & \text{otherwise}\\
\end{cases},\\
\end{equation}

\begin{equation}\label{eq5}
OP_{l,t}=\begin{cases}
O_{p,j}, &\text{if $X_{p,j,k,l,t}=1$}\\
OP_{l,t-1}, & \text{otherwise}\\
\end{cases},\\
\end{equation}

\begin{align}\label{eq6}
C_{p,j,k}=\sum_{l\in\mathbf{M}_{p,j}}\sum_{t=1}^{NS}&(Z_{p,j,l,t}X_{p,j,k,l,t}(t+PR_{p,j}U_{p,k})+(1-Z_{p,j,l,t})X_{p,j,k,l,t} \nonumber\\
&(t+{CO}_{{PT}_{l,t},{OP}_{l,t},p,O_{p,j}}+P_{p,j}U_{p,k}))\: \forall p,j,k,
\end{align}

\begin{align}\label{eq7}
C_{p,j+1,k}\geq &C_{p,j,k}+PR_{p,j+1}U_{p,k}+\sum_{l\in\mathbf{M}_{p,j+1}}\sum_{t=C_{p,j,k}}^{NS} \nonumber \\ &(1-Z_{p,j+1,l,t})X_{p,j+1,k,l,t}{CO}_{{PT}_{l,t},{OP}_{l,t},p,O_{p,j+1}}\: \forall p,j,k,
\end{align}

\begin{equation}\label{eq8}
C_{p,j,k} - PR_{p,j}U_{p,k}\geq C_{p,j,k-1} - PR_{p,j}U_{p,k-1} \:\forall p,j,k,
\end{equation}

\begin{equation}\label{eq9}
\sum_{t=1}^{NS}\sum_{l\in\mathbf{M}_{p,j}} X_{p,j,k,l,t}\le1\: \forall p,j,k,
\end{equation}

\begin{align}\label{eq11}
&\left(TH-\sum_{p}\sum_{k=1}^{K_p}\sum_{j=1}^{J_p}\sum_{\tau=\left(n-1\right)S+1}^{t}{(1-Z_{p,j,l,\tau})}X_{p,j,k,l,\tau}{CO}_{{PT}_{l,t},{OP}_{l,t},p,O_{p,j}}\right)\sum_{p}\sum_{k=1}^{K_p}\sum_{j=1}^{J_p}\nonumber\\&\sum_{\tau=t}^{nS}(1-Z_{p,j,l,\tau})X_{p,j,k,l,\tau}\geq0 \:\forall l,t\in\{\left(n-1\right)S+1,\ldots,nS\},n\in\{1,\ 2,\ldots,N\}.
\end{align}

Equations (\ref{eq2}) and (\ref{eq3}) make sure that a machine cannot process multiple jobs simultaneously. Additionally, (\ref{eq4}) and (\ref{eq5}) ensure that a machine only changes its setup when necessary, such as when a new product type or operation is assigned. If a setup change occurs, the machine requires both processing time and conversion time to complete the job, but only processing time is required if there is no setup change, as stipulated in constraint (\ref{eq6}). Here, $NS$ represents the total number of decision points occurring within an episode, which spans the planning horizon of $N$ shifts. To ensure that jobs are executed in the proper sequence, (\ref{eq7}) guarantees that an operation can only be executed once the previous operation for a job is completed. The first-in, first-out (FIFO) rule is applied to lots of the same product type that are waiting for the same operation, as described in constraint (\ref{eq8}). Equation (\ref{eq9}) guarantees that a lot can only be assigned to one machine once for an operation. If the lot requires reentry into an operation, that reentry is considered a separate operation in sequence $\{O_{p,j}|j=1, 2, \cdots, J_p\}$. In this way, we can incorporate the reentry feature of factory scheduling into our model. Finally, constraint (\ref{eq11}) ensures that a conversion cannot be executed if the cumulative conversion time exceeds $TH$ within a single shift.

\section{Scheduling model for factory-wide DFJSSP} \label{sec:model}
In this section, we introduce a scheduling model based on MARL designed to effectively address the factory-wide DFJSSP. The proposed model features two types of agents: followers and a leader. We begin by presenting the formulation for the followers and the leader, detailing their respective state and action representations along with the reward mechanisms. Next, we describe the simulation environment used for model training and training algorithm employed to train the agents within our scheduling model. Finally, to minimize production capacity loss by preventing catastrophic mistakes by agents, we introduce a rule-based conversion algorithm.

\subsection{Follower model}
\begin{figure}[h]
\centering
\subfloat[]{\includegraphics[width=\linewidth]{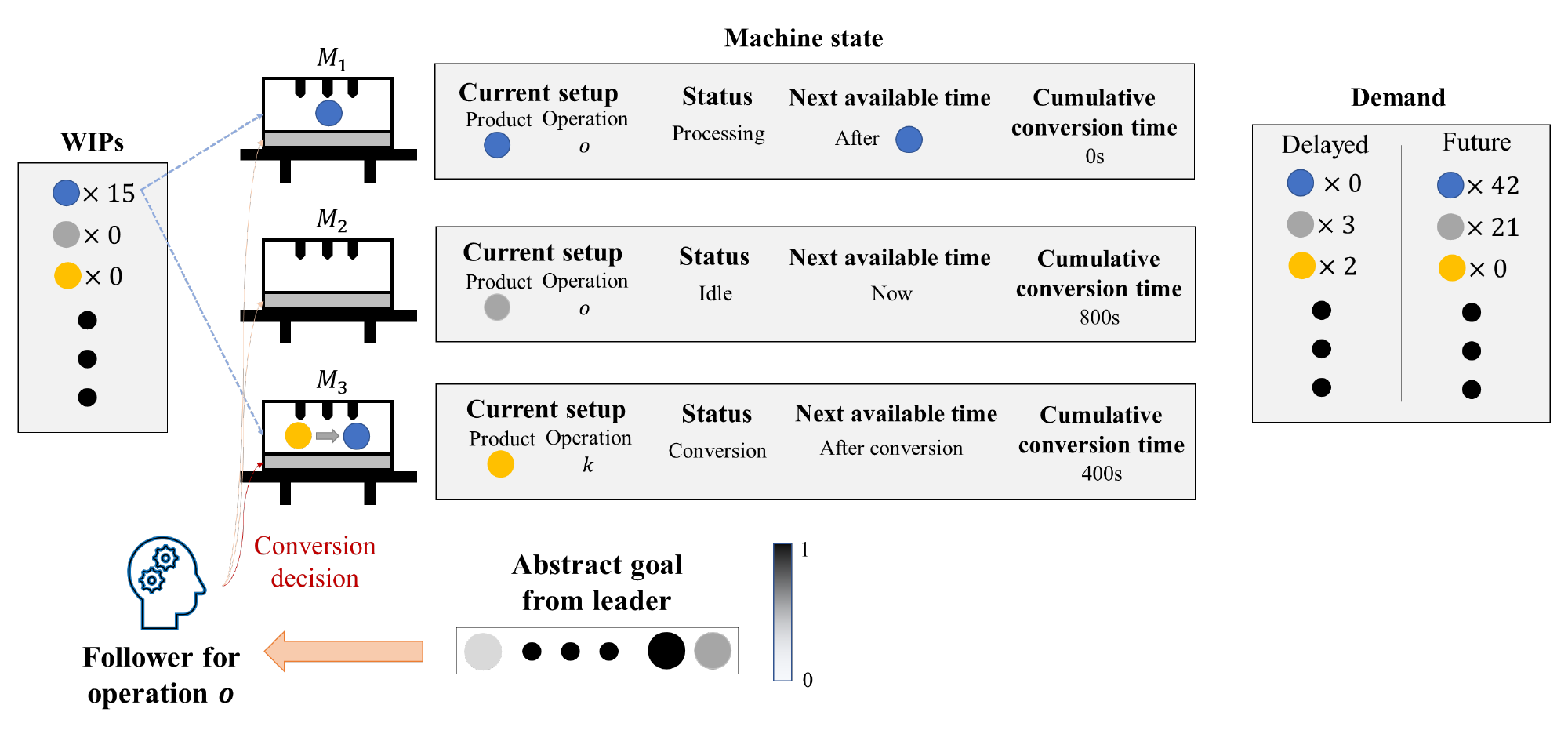}%
\label{fig_1a}}
\par
\subfloat[]{\includegraphics[width=0.9\linewidth]{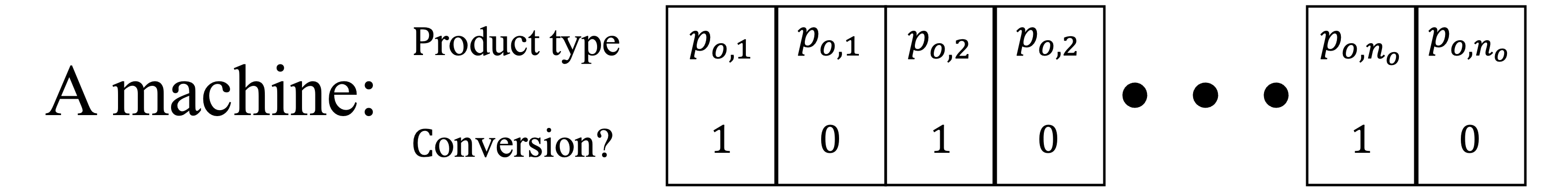}%
\label{fig_1b}}
\caption{Illustration of (a) state and (b) action of the follower for operation $o$.}
\label{fig_1}
\end{figure}

Our proposed model designates each follower (i.e., a lower-level agent) the responsibility of managing its own operation. As depicted in Fig. \ref{fig_1a}, the state of each follower includes information about the state of machines, work-in-process (WIP), and demand. Specifically, the machine state elements, such as the current setup ($PT_{l,t}$ and $OP_{l,t}$) and status ($Q_{l,t}$), are encoded using one-hot encoding to represent the current product and operation for a machine $l$ at time $t$. Both scheduled and unscheduled maintenance are represented identically by setting $Q_{l,t}=1$. The cumulative conversion time for the current shift and the next available time are also tracked to predict future availability of the machine and products. The cumulative conversion time of machine $l$ at $t$ is calculated based on $\{PT_{l,t_0}, OP_{l,t_0}, CO_{p,o,PT_{l,t_0},OP_{l,t_0}}| \forall o, p, t_0\leq t\}$, and the next available time is determined from $\{U_{p,k}, PR_{p,j}, X_{p,j,k,l,t_0} | \forall p, j, k, t_0\leq t\}$. If scheduled maintenance is in progress, the next available time is set to the end of the maintenance period. Each agent maintains these machine state elements for all machines pertinent to the target operation. Additionally, WIPs awaiting the operation and both delayed and future demand lots for all available product types in the target operation are encapsulated in the state vector. These can be readily derived from $\{D_{p,k}, C_{p, j, k}, X_{p,j,k,l,t_0}| \forall j, k, l, t_0\leq t\}$ for product type $p$. The underlying formulas are given in (\ref{eq2})-(\ref{eq11}). Additionally, at the beginning of each shift, the leader distributes abstract goal vectors for each follower, and the abstract goal is incorporated into the follower's state to direct them towards achieving a higher reward. Details are provided later.

Product setup changes occur intermittently in real-world factories. Consequently, the followers must make crucial decisions regarding conversion for all machines assigned to the operation, as illustrated in Fig. \ref{fig_1b}. To accomplish this, followers produce action vectors $\bm{a}^o_{n,t}$ for all machines within the target operation $o$ at $t$ in the $n$-th shift. If multiple machines belong to operation $o$, the follower $o$'s action is the concatenation of actions for all member machines. That is, $\bm{a}^o_{n,t}=(\bm{a}^{o,l}_{n,t}|l \in \text{available machines for operation } o)$, where $\bm{a}^{o,l}_{n,t}$ is the action for the available machine $l$ in operation $o$. Each action vector $\bm{a}^{o,l}_{n,t}$, a one-hot encoded vector, represents two factors for a machine: whether or not to convert and the next product type. In Fig. \ref{fig_1b}, $p_{o,i}$ is $i$-th available product type in operation $o$ and $np_o$ is the number of available product types in operation $o$. 
If the first entry is activated, $p_{o1}$ is selected as the next product type. Otherwise, there is no change made to the product setup. Let $\psi_{l,t}^{p_{o,i},1}$ be set to 1 if $p_{o,i}$ is chosen as the next product setup for machine $l$ at $t$ via the activation of $(\bm{a}^{o,l}_{n,t})_{p_{o,i},1}$. Conversely, $\psi_{l,t}^{p_{o,i},0}$ should be set to 1 either if $(\bm{a}^{o,l}_{n,t})_{p_{o,i},0}$ is activated or if another product is chosen as the setup. By default, both $\psi_{l,t}^{p_{o,i},1}$ and $\psi_{l,t}^{p_{o,i},0}$ are initialized to 0. Then, for any machine $l$, we have
\begin{equation}\label{setup_constrain1}
\psi_{l,t}^{p_{o,i},0}+\psi_{l,t}^{p_{o,i},1}=1
\end{equation}
and
\begin{equation}\label{setup_constrain2}
\sum_{i = 1}^{np_o}\psi_{l,t}^{p_{o,i},1} \leq 1.
\end{equation}

The next product type is only relevant when a conversion is triggered, and conversions can only occur if there is WIP of the selected product in the operation. Let us assume that $p_{o,i}$ is selected and a conversion is initiated by an action vector for machine $l$ at time $t$, corresponding to operation $o = O_{p,j}$. If there is a waiting lot $k$, then $X_{p_{o,i},j,k,l,t+1}$ is set to 1; otherwise, it remains 0. Conversely, if no conversion is triggered but there is an upcoming lot $k$ for the current setup $p_{o,i}$ of machine $l$ at time $t$, then $X_{p_{o,i},j,k,l,t+1}$ is also set to 1.

It is not always necessary for all followers to take action since there may be a lack of available machines, as depicted in Fig. \mbox{\ref{fig_2b}}. In this situation, only the followers who have access to at least one available machine are able to produce actions. The action of the follower $o$ at $t$ in the $n$-th shift, $\bm{a}^o_{n,t}$, is determined according to
\begin{equation}\label{eq_pi_i}
\bm{a}^o_{n,t} \sim \pi^{o}\left(\;\cdot\;\vert \bm{s}^o_{n,t}, \bm{g}^o_n\right),
\end{equation}
where $\pi^{o}$ represents the policy of the follower responsible for operation $o$, while $\bm{g}^o_n$ denotes the goal vector for the follower during the $n$-th shift. 

In the MARL-SR algorithm \mbox{\cite{jang2023learning}}, synthetic rewards are utilized to account for each agent's contribution to the overall team reward. However, given that the demand for each product type is specified for each shift, we apply a more straightforward reward calculation method. At the end of each shift $n$, followers are penalized based on the number of delayed lots. Specifically, for the operation $o$, the reward $r^{o}_t$ is defined as follows:
\begin{equation}\label{operation_reward}
r^{o}_{n}=-\sum_{i = 1}^{np_o}\sum_{k=1}^{K_{p_{o,i}}}H_{p_{o,i},j, k},\:\text{where }\: H_{p_{o,i},j, k}=\begin{cases}
1, &\text{if $D_{p_{o,i},k}<C_{p_{o,i},j,k}$ and $D_{p_{o,i},k} \leq nS$}\\
0, & \text{otherwise}.\\
\end{cases}\\
\end{equation}
This operational rewarding strategy is the default approach for our followers. The training environment is quite challenging, with the majority of actions being trivial and sparse rewards being provided. To supplement potentially insufficient training and prevent followers from generating inadequate actions that can result in catastrophic production capacity losses, we introduce a rule-based conversion algorithm in Section \ref{sec:rule-based conversion}.

\subsection{Leader model}
Factory-wide DFJSSP requires a high level of coordination between multiple entities, such as jobs, machines, and operations. Traditional single agent based RL methods do not perform well due to the exponential growth of the joint action space as the number of entities increases \cite{Rashid2018}. To address this challenge, MARL has been proposed in which each agent aims to obtain the best policy for an entity or sub-problem \cite{Zhang2021a}. This decentralization is based on the premise that achieving individual goals assigned to each agent is easier than attaining team success. However, individual success is meaningless unless team goals are achieved. Therefore, it is crucial to maximize team reward through cooperation while pursuing individual goals in MARL.

One approach of MARL is to train decentralized policies in a centralized manner, as this approach can offer valuable state information and eliminate communication barriers among agents \cite{Rashid2018, Yang2020, Oliehoek2007, Kraemer2016, Lowe2017}. Despite the increasing interest in centralized training with decentralized execution in the RL community, this method is not viable for use in advanced production systems due to the high synchronization demand between agents. Especially, factory-wide scheduling involves numerous precedence constraints that can be represented as a DAG. Consequently, this scheduling problem can be modeled as a multi-agent problem with DAG constraints, where each agent handles a specific operation. Essentially, every agent aims to maximize team rewards, considering the complex interrelationships between agents. A significant challenge is that rewards can only be ascertained after multiple agents have made a series of sequential decisions over an extended period. This delay in reward distribution complicates an agent's ability to assess its contribution to team rewards at each time step, thus impeding effective learning from the perspective of the individual agent. To overcome these challenges, we use the concept of the leader proposed in \cite{jang2023learning}. This outer agent aims to coordinate the operation-specific agents towards achieving high team rewards by setting long-term, agent-specific goals. 

\begin{figure}[h]
\centering
\subfloat[At the beginning of $n$-th shift]{\includegraphics[width=0.47\linewidth]{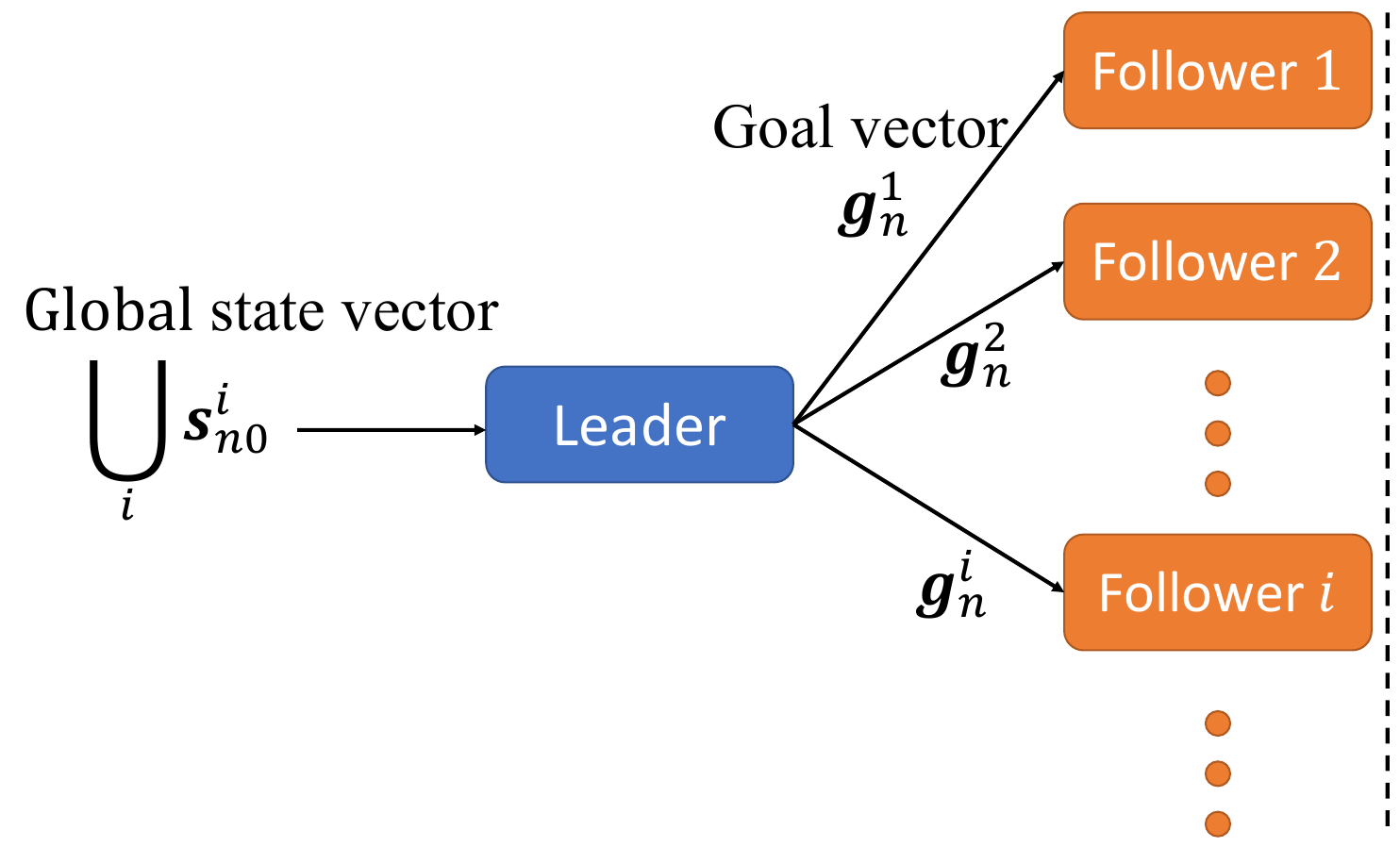}%
\label{fig_2a}}
\subfloat[At $t$ in $n$-th shift]{\includegraphics[width=0.51\linewidth]{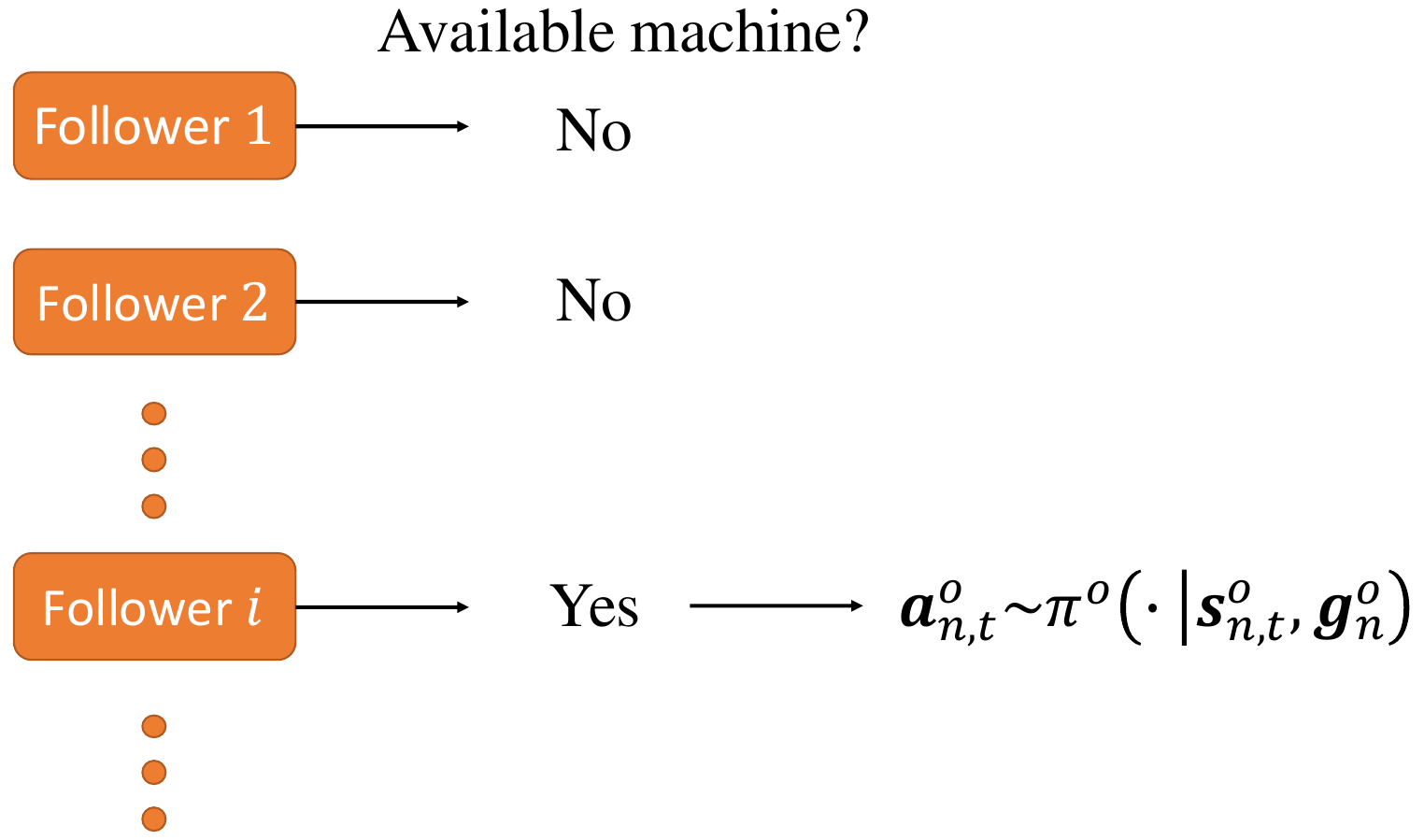}%
\label{fig_2b}}
\caption{Overview of our MARL scheduling algorithm based on (a) leader and (b) followers. Here, $\pi^o$ is the incumbent policy of the follower.}
\label{fig_2}
\end{figure}

The leader generates goals by taking into account all relevant factors, including the states of operations, and distributes them to each follower at the beginning of every shift. These goals are expressed in an abstract form as real numbers in [0, 1] and serve as communication channels between the leader and followers. The leader's rewards are based on the followers' achievements at the end of each shift. Therefore, the leader needs to create goal vectors that contain meaningful messages to lead the followers towards higher team rewards by observing the global environment information. The team rewards are also given to the leader. The role of the leader is illustrated in Fig. \ref{fig_2a}. At the beginning of each shift, the leader receives the global state vector $\cup_o \bm{s}^o_{n,0}$, where $\bm{s}^o_{n,0}$ is the initial state vector of operation $o$ in the $n$-th shift. The leader utilizes the global state vector to generate individualized goals for each follower, which are then used to address specific operations. The leader's policy $\pi^{L}$ is defined as
\begin{equation}\label{eq_pi_m}
(\bm{g}^o_n|o \in \text{all operations}) \sim \pi^{L}\left(\;\cdot\;\vert\bigcup_o \bm{s}^o_{n,0}\right).
\end{equation}
This approach makes it much easier to handle complex factory-wide scheduling problems, as they are decomposed into a set of weakly connected sub-scheduling problems and the problem of managing them.

While the leader has access to comprehensive information about the team's operations, followers only possess partial information. Nonetheless, they can infer the overall state of the system, including the state of the other operations and the coordination strategy of the leader, from abstract goals. Therefore, establishing effective communication channels is crucial for attaining high team rewards, utilizing goal vectors and team rewards.

The primary objective of scheduling in this study is to maximize completion rate, and the reward for maximizing completion rate can take the form of the negative value of the number of lots as introduced in Section \ref{sec: PF}. In addition, when creating schedules, it is assumed that the demand for each product type is given per shift. Thus, we calculate the penalty based on delayed lots in each operation after each shift. Then, a follower receives the sum of the penalties for all processable product types in the assigned operation as the reward after each shift based on (\ref{operation_reward}). For the leader, who is responsible for overseeing the status of all operations, all delayed lots are considered when calculating the reward and it is given to the leader after each shift $n$ as follows:
\begin{equation}\label{leader_reward}
r^{l}_{n}=-\sum_{p}\sum_{k=1}^{K_p}H_{p, J_p, k},\:\text{where}\: H_{p, J_p, k}=\begin{cases}
1, &\text{if $D_{p,k}<C_{p,J_p,k}$ and $D_{p,k} \leq nS$}\\
0, & \text{otherwise}.\\
\end{cases}\\
\end{equation}

\subsection{Environment and Training}
In this section, we describe the simulation environment used to train RL agents, as summarized in Algorithm \ref{alg_environment}. Initially, we configure the production environment details, including the planning horizon ($NS$), product types ($p$), sequences of operations for each product type ($O_{p,j} \forall p, j \in 1,2,\cdots, J_p$), sets of compatible machines ($\bm{M}_{p,j} \forall p, j \in 1,2,\cdots, J_p$), unit processing times ($PR_{p,j} \forall p, j$ $\in 1,2,\cdots, J_p$), scheduled maintenance details ($SM_{l,t} \forall l, t$), and the conversion time threshold ($TH$). Each episode involves generating production plans for each shift over a predefined duration, detailing the required number of lots for each product type based on the incumbent policy. For each episode, it is assumed that demand information, specifying the required number of lots for each product type per shift, is provided. The objective of the RL agents in each episode is to meet this demand without delay. This demand information serves as the basis for reward calculation. Historical data contains only few demand observations; definitely insufficient for RL training. For this reason, we randomly generate demand for each episode with the underlying distribution calibrated from historical data, ensuring it accurately reflects the actual factory environment. We consider various factors such as the distribution of demand by product type, along with a range of operational constraints and conditions - including precedence constraints, processing times, and conversion times by product type - all based on real production datasets from a high-volume packaging and test factory. Furthermore, the initial settings of the factory environment, including the status of machines and WIP, are configured at the beginning of each episode. These initial settings lay the foundation for generating schedules based on the policies of the leader and the followers.

As is typical in RL, given an incumbent policy, at the beginning of the first shift of each episode, the leader creates and distributes goal vectors, as defined in (\mbox{\ref{eq_pi_m}}), which guide the followers towards achieving high team rewards. At each decision point, machine statuses are updated to `unavailable' based on both scheduled ($SM_{l,t} \forall l$) and unscheduled maintenance information ($\bm{\Omega}_t$), which are not considered in the MARL-SR algorithm \mbox{\cite{jang2023learning}}. The distribution of the latter is generated based on historical downtimes that depend on the machine type and the specific operations performed by each machine. A decision can only be made if a machine is available, meaning it is not processing, undergoing a setup change, or undergoing scheduled or unscheduled maintenance; otherwise, no action is taken. Subsequently, the followers make conversion decisions or maintain the current product type at every decision point, including the initial time of each shift, based on (\mbox{\ref{eq_pi_i}}) and a rule-based conversion algorithm (detailed in Section \mbox{\ref{sec:rule-based conversion}}). Once all decisions by available followers are made, the statuses of the machines and WIP information are updated. At the end of each shift, rewards are assigned to each follower and the leader as specified in (\ref{operation_reward}) and (\ref{leader_reward}) respectively; however, no rewards are distributed during a shift. This is because the designated due time for meeting demand coincides with the end of the shift, resulting in a reward value of zero during the shift. As a result, after each shift, multiple ($s, a, r$) triplets corresponding to the number of decisions made by each follower are sampled, along with a single triplet for the leader. Each episode is stochastic with respect to the policy, demand, and unscheduled maintenance.

\begin{algorithm} 
 \caption{A summary of simulation environment}
\label{alg_environment}
 \begin{algorithmic}[1]
\REQUIRE Episode duration $NS$, all producible product types $p$, sequences of operations for each product type $O_{p,j}, \forall p, j \in 1,2,\cdots, J_p$, sets of compatible machines $\bm{M}_{p,j}, \forall p, j \in 1,2,\cdots, J_p$, unit processing times $PR_{p,j}, \forall p, j \in 1,2,\cdots, J_p$, scheduled maintenance information $SM_{l,t} \forall l, t$, and the conversion time threshold $TH$
\LOOP
\STATE Configure the initial setup of each episode using job information ($D_{p, k} \text{ and } U_{p,k} \forall p, k \in 1,2,\cdots,K_p$), as well as the initial settings for machines ($Q_{l,0}, PT_{l, 0}, \text{ and } OP_{l,0} \forall l \in \bm{M}$)
\STATE Sample random demand
\STATE The initial state vectors of the leader and the followers are set
\FOR {each shift}
\FOR {each decision point}
\STATE Sample unscheduled maintenance
\IF {initial decision point}
\STATE The leader creates and distributes goal vectors based on (\ref{eq_pi_m})
\ENDIF
\STATE Machine statuses are updated to account for both scheduled and unscheduled maintenance based on (\ref{eq2})
\STATE All available followers take actions based on (\ref{eq_pi_i}) and the rule-based conversion algorithm in Section \ref{sec:rule-based conversion}
\STATE Machine statuses are updated to account for the actions of the followers based on (\ref{eq2})
\IF {the end of the shift}
\STATE Rewards are assigned to each follower and the leader as specified in (\ref{operation_reward}) and (\ref{leader_reward}), respectively
\STATE The state vector of the leader is updated
\ELSE
\STATE Each follower receives a reward value of zero.
\ENDIF
\STATE The state vectors of the followers are updated
\ENDFOR
\ENDFOR
\ENDLOOP
\end{algorithmic}
\end{algorithm}



Although our goal is to enhance coordination through the leader, each agent functions individually, communicating via the goal vector. Consequently, samples are collected separately for each agent, as detailed in Algorithm \ref{alg_environment}. Therefore, we train each agent using their own rollouts. Specifically, we employ the proximal policy optimization (PPO) algorithm \cite{Schulman2017} as the baseline algorithm to train all agents. PPO has received huge attention due to its simplicity, robustness for a large variety of tasks, low computational complexity, and good performance \cite{Zhen2020, Zhang2019c}.

\subsection{Rule-based conversion algorithm} \label{sec:rule-based conversion}
Although a follower only needs to consider a single operation, there are usually numerous product candidates to choose from during a conversion. However, executing a conversion to start a product type can result in a significant loss of capacity, as it may take away the chance of producing other products due to the limited conversion time. To mitigate the risk of such a loss, we propose a rule-based conversion algorithm.

To determine the appropriate product type based on the rule-based conversion algorithm, we first score the urgency of product types for each operation using Algorithm \ref{alg_1}. We begin by calculating the required capacity and expected remaining capacity for each product type. For a given product type $p$ and operation $o$, the required capacity ($RC_{o,p}$) is the estimated total processing time for all target product WIPs in the target operation. The expected remaining capacity ($ERC_{o,p}$) is the estimated total processable time until the planning horizon ends of all machines that are currently processing $p$ in operation $o$. If the required capacity is greater than the expected remaining capacity, we consider that product to be urgent and set the urgency score to the required capacity. Furthermore, a product that is not currently being processed by any machine in the operation is considered much more urgent because it cannot be produced without a conversion. In this case, we add a large value ($BN$) to the urgency score to prioritize this product. It is important to note that $BN$ must be greater than the maximum value of the possible required capacity. The urgency score of $o$ and $p$ is denoted by $US_{o,p}$

\begin{algorithm} 
 \caption{Urgency scoring}
\label{alg_1}
 \begin{algorithmic}[1]
\REQUIRE Required capacity $RC_{o,p}$ and expected remaining capacity $ERC_{o,p}$ for all pairs of (operation $o$, product $p$)
\STATE Set a large value $BN$ for urgency scoring
\STATE Initialize urgency score $US_{o,p}=0$ for all pairs of (operation $o$, product $p$)
\FOR {each operation $o$}
\FOR {each product type $p$ processable in operation $o$}
\IF {$RC_{o,p}>ERC_{o,p}$}
\IF {$p$ is not being processed by any machine in operation $o$}
\STATE $US_{o,p}=RC_{o,p} + BN$
\ELSE
\STATE $US_{o,p}=RC_{o,p}$
\ENDIF
\ENDIF
\ENDFOR
\ENDFOR
\end{algorithmic}
\end{algorithm}

After urgency scoring computation in Algorithm \ref{alg_1}, we make a conversion decision for each available machine as outlined in Algorithm \mbox{\ref{alg_2}}. We first create an action for each operation $o$ according to policy (\mbox{\ref{eq_pi_i}}). Subsequently, we decode the corresponding action for each available machine to determine two key variables: the conversion indicator ($CI$), which indicates whether a conversion should be triggered, and the candidate product ($p_{cand}$), which represents the potential next product setup. When determining the next product for an available machine $m$, we take into account not only the generated action for that machine but also the current product WIPs and the operation's capacity for that current product. If there are WIPs for the selected product type, we assign $p_{cand}$ to the next product setup in the episode. Otherwise, we prioritize urgent products and select the most urgent product as the next product setup. To accomplish this, we evaluate the current product setup $p_m$ on machine $m$ and determine if it can be replaced with the most urgent product, $\argmax_{p \in P_o}US_{o,p}$, where $P_o$ is the set of eligible products by operation $o$. To ensure that there will be no missing lots of $p_m$, we first verify that all WIPs of $p_m$ can be completed before the planning horizon ends. We do this by verifying that the required capacity $RC_{o,p_m}$ is less than $ERC_{o,p_m}-ERCM_{o,m,p_m}$, the expected remaining capacity without the target machine $m$. Here, $ERCM_{o,m,p_m}$ refers to the expected remaining capacity of the target machine $m$. Additionally, we check if there is an urgent product ($\sum_{p \in P_o}US_{o,p}>0$). If both conditions are met, we assign the most urgent product as the next product for that machine.

\begin{algorithm} 
 \caption{Conversion decision at $t$ in shift $n$}
\label{alg_2}
 \begin{algorithmic}[1]
\REQUIRE Required capacity $RC_{o,p}$ and expected remaining capacity of operation $ERC_{o,p}$ for all pairs of (operation $o$, product $p$), expected remaining capacity of machine $ERCM_{o,m,p}$ for all triplets of (operation $o$, machine $m$, product $p$)
\STATE \textit{Urgency scoring}()
\FOR {each operation $o$}
\IF {there is an available machine}
\STATE Produce $\bm{a}^o_{n,t}=\{\bm{a}^{o,m}_{n,t}|m \in \text{machines for operation } o\}$
\FOR {each available machine $m$}
\STATE Get current product setup $p_m$
\STATE $CI$ and $p_{cand}$ $\leftarrow$ \textit{Decode} ($\bm{a}^{o,m}_{n,t}$)
\IF {$CI$=True}
\IF {there are WIPs of $p_{cand}$ in operation $o$}
\STATE Next product setup $p_m$ $\leftarrow$ $p_{cand}$
\ELSIF{$RC_{o,p_m}<ERC_{o,p_m}-ERCM_{o,m,p_m}$ and $\sum_{p \in P_o}US_{o,p}>0$}
\STATE Next product setup $p_m$ $\leftarrow$ $\argmax_{p \in P_o}US_{o,p}$
\ELSE
\STATE Do not change product setup
\ENDIF
\ENDIF
\ENDFOR
\ENDIF
\ENDFOR
\end{algorithmic}
\end{algorithm}

In summary, we introduce several key modifications to the MARL-SR algorithm outlined in \mbox{\cite{jang2023learning}}:
\begin{itemize}
\item MARL-SR employs synthetic rewards to distinguish and reflect each agent's contribution to the overall team reward, as individual agent reward calculation is typically infeasible. In contrast, for factory scheduling, we can measure operation-wise performance, allowing us to implement an operation-specific reward calculation.
\item Additionally, our approach models both scheduled and unscheduled machine maintenance, which are not considered in MARL-SR.
\item Given that poor decisions by an agent can severely impact production capacity in factory scheduling, we also develop a rule-based conversion algorithm to override agent decisions if they could lead to significant production losses.
\end{itemize}

\section{Experimental Evaluations}\label{sec:experiment}

\subsection{Implementation details}
We implemented both the proposed model and the benchmark models using the PPO algorithm. For each agent, a fully-connected neural network, consisting of two layers with 256 units each and ReLU activation functions, is utilized for both the actor and the critic components. In our algorithm, the actor consists of the leader and the followers which all have such network architecture. We compared our proposed model with two benchmark models, which are described in detail later in this section. To determine the optimal hyperparameters for each model, we conducted both Kruskal-Wallis and ANOVA tests for reasonable parameter choices and groups corresponding to performance metrics. The results showed no significant differences across the four performance metrics-tardiness, number of changeovers, cumulative idle time, and completion rate against demand-within the parameter ranges tested: batch size (64-256), learning rate (0.0001-0.00001), discount factor (0.95-0.99), and lambda (0.9-0.95) for our proposed model and one of the benchmark models. For the other benchmark model, performance improvement was observed only when lambda was set to 0.95, with no statistical differences noted for the other hyperparameters. Consequently, we applied the hyperparameter settings listed in Table \mbox{\ref{table_hi}} to all models. Additionally, for each operation $o$ in shift $n$, the dimension of the corresponding goal vector $g^o_n$ is set to 3, as recommended in \mbox{\cite{jang2023learning}}. All models compared in this study are implemented using TensorFlow version 2.8 on a server equipped with an Intel i9-12900k CPU. Due to the shallow network architecture employed for both the actor and the critics, and the simulation's need for extensive sequential computations, GPU execution does not help.

\begin{table}[h]
\caption{Hyperparameters}
\label{table_hi}
\centering
\begin{tabular}{l|r}
\hline
\multicolumn{1}{c|}{Hyperparameter} & \multicolumn{1}{c}{Value}\\
\hline
Batch size & 256 \\
Learning rate & 0.0001 \\
Discount factor & 0.9900 \\
Clipping value & 0.2000 \\
Generalized advantage estimation parameter lambda & 0.9500\\
\hline
\end{tabular}
\end{table}

In this study, we evaluate the effectiveness of our proposed method in real-world scheduling scenarios using two simulators built on different real production datasets for Intel’s high volume packaging and test factory. We refer to the two production scenarios as short-term and long-term scenarios, as summarized in Table \ref{table_1}. Here, the average out-degree in DAG refers to the mean number of outgoing arcs in the DAG, which is constructed based on the precedence constraints of operations. To assess performance in diverse environments, we further diversified each scenario into low-demand, medium-demand, and high-demand cases. Specifically, compared to low-demand cases, medium- and high-demand cases require three and five times the demand on average, respectively. In general, a more sophisticated scheduling strategy is required as demand increases. 

To enhance exploration during training, the set of target product types required for production in each episode changes probabilistically, resulting in varying workloads for each agent. Additionally, to promote exploration within the PPO algorithm, we incorporate entropy regularization as outlined in \mbox{\cite{SchulmanAC17}}, with the entropy coefficient of 0.01. This regularization helps maintain policy diversity by preventing premature convergence and fostering a broader exploration of the action space.

For each demand case in both short- and long-term scenarios, we train the comparison models over 10,000 episodes, with each model utilizing the policies that achieved the highest team reward during the last 100 episodes of training. These optimal models are then tested over 100 episodes. In each test episode, all comparison models begin scheduling under identical conditions, including the same demand information as well as both scheduled and unscheduled maintenance details.

In our algorithm, the followers learn policies to meet the demand over a predefined number of shifts. However, the number of shifts in the two cases in Table \mbox{\ref{table_1}} is too long to enable effective learning within a reasonable timeframe. Therefore, we train the agents in simulations spanning four shifts (two days) and then apply the learned policies using a rolling horizon approach during inference. In the long-term scenario, training takes 5.62 hours, with inference requiring 2.68 minutes per episode. For the short-term scenario, training takes 5.92 hours, and inference takes 0.53 minutes per episode. Given the planning horizon for both scenarios, the time required for both training and inference is manageable.


\begin{table}[h]
\caption{Summary of two scenarios}
\label{table_1}
\centering
\begin{tabular}{l|c|c}
\hline\hline
Scenario & Short-term & Long-term \\
\hline 
No. product types & 35 & 20 \\
No. operations & 26 & 21\\
No. machines & 159 & 115\\
Avg. out-degree in DAG & 1.33 & 1.37\\
\multirow{2}{*}{Duration} & 1 week  & 3 weeks\\
 & (14 shifts)  & (42 shifts)\\
\hline\hline
\end{tabular}
\end{table}

In this study, we compare the proposed model with existing RL-based dynamic scheduling models. While several such models have been developed, most are not suitable for factory-wide scheduling because existing models do not consider factory-specific operational constraints in modeling or require excessive computational time. Two recent models - deep reinforcement learning for job shop scheduling problems (DRL-JSSP) \cite{Liu2020} and deep reinforcement learning for dynamic flexible job shop scheduling (DRL-DFJSS) \cite{Chang2022} - can be adapted for factory-wide scheduling with some modifications. Both models use RL agents to select dispatching rules at each action point. DRL-JSSP employs a single agent per operation to choose one of the three rules, while DRL-DFJSS trains one agent to handle all operations and selects one of the four rules. However, as the third rule corresponding to DRL-DFJSS is not applicable in our problem, we excluded it from the analysis. 

In a factory environment, the large state and action vectors necessary to manage hundreds of machines lead to significant learning and inference times. Moreover, the state representation and reward calculation methods used in DRL-JSSP and DRL-DFJSS are not directly applicable to our simulation, where the number of jobs demanded varies and the penalties for tardy jobs are unknown. Therefore, we employ the same multi-agent setting where an agent is assigned to each operation (with the concept of a leader being unique to our model), use a consistent state representation feature format (with our model incorporating goal vectors from the leader), and apply an operation-wise rewarding strategy across all models, focusing on comparing the algorithmic differences in job selection and agent coordination.

\subsection{Comparison with benchmarks} \label{comp:sota}
\begin{figure}[h]\centering
  \includegraphics[width=0.6\textwidth]{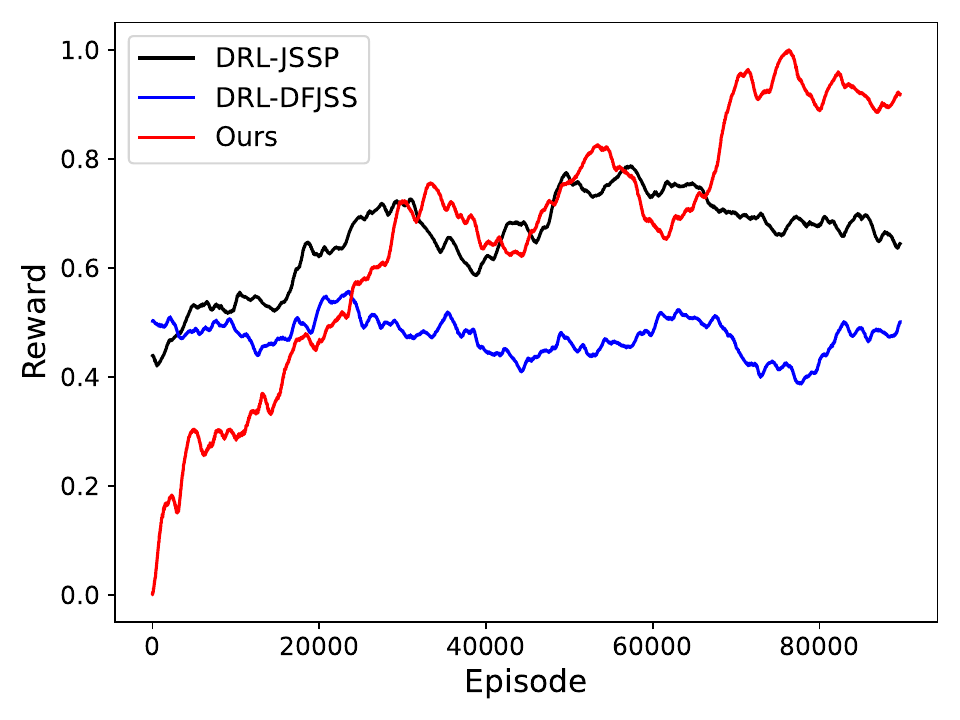}
  \caption{Training of the three models on the low-demand case of the long-term production scenario. Min-max normalization is applied to the total reward to standardize the scale of the y-axis.}
  \label{fig_3}
\end{figure}

In this section, we compare the proposed model with existing RL-based dynamic scheduling models. Fig. \ref{fig_3} illustrates the total reward per episode for our proposed model and the two comparison models trained on the low-demand case of the long-term production scenario. In the figure, the moving window method is applied, where the team rewards are averaged over a sliding window of 10,000 episodes. The averaging window is shifted by one episode at a time, indicating a step size of one episode per move. The DRL-JSSP and DRL-DFJSS models initially perform well due to their reliance on human-designed dispatching rules. However, their performance remains almost stagnant throughout training, indicating that a combination of dispatching rules is insufficient for large-scale factory-wide scheduling. Conversely, the proposed model exhibits poor performance initially but demonstrates a substantial improvement during training. This implies that the proposed model can effectively address the challenges of large-scale factory-wide scheduling.

Next, we utilize four metrics, namely tardiness, the number of changeovers, cumulative idle time, and completion rate against demand, to compare the models. We are unable to divulge the performance of the proposed model due to Intel's data confidentiality policy. Consequently, we establish the proposed model as the baseline and assess the improvement (percentage improvement) over benchmark models. Specifically, we report an increase in completion rate and a decrease in tardiness, the number of changeovers, and cumulative idle time as performance improvement. In summary, a negative number means that we outperform.

Tables \ref{table_3} and \ref{table_4} show comparison results for the long- and short-term production scenarios, respectively. We highlight the improvement values of the comparison models in bold if the proposed model achieves the best performance for the corresponding criterion. In summary, the proposed model achieves the best performance for 10 out of 12 criteria in both long- and short-term production scenarios. Moreover, the performance gap generally increases as the demand level rises, except in the case of the number of conversions. Notably, our proposed model consistently delivers the best performance in terms of conversions, despite this metric's apparent lack of correlation with demand levels. Although the proposed model shows a significant decline in cumulative idle time for the low-demand case of the long-term scenario, it excels in more critical metrics such as tardiness and completion rate, which are directly tied to productivity. Furthermore, despite a noticeable drop in tardiness performance during low-demand conditions in the short-term scenario, the gap is relatively minor. This is because all three models maintain very low tardiness when demand is low, as reflected by the standard deviation in percentage improvement.

\begin{table}[h]
\caption{Percentage improvement of comparison models in four metrics compared to the proposed model for the long-term production scenario}
\label{table_3}
\centering
\begin{tabular}{c|c|r|r|r|r}
\hline\hline
 \multirow{2}{*}{Demand} & \multirow{2}{*}{Metric} & \multicolumn{2}{c|}{DRL-JSSP} & \multicolumn{2}{c}{DRL-DFJSS}\\
 \cline{3-6}
 &  & Mean & Std. & Mean & Std.\\
 \hline
\multirow{4}{*}{Low}& Tardiness & \textbf{-18.91}&106.65 & \textbf{-10.06}&111.45 \\
 & No. conversions & \textbf{-121.50}&57.07 & \textbf{-139.91}&66.26 \\
 & Cumulative idle time & 17.27&7.19 & 16.51&7.88 \\
 & Completion rate & \textbf{-1.25}&4.51 & \textbf{-0.83}&4.38 \\
\hline
\multirow{4}{*}{Medium}& Tardiness & \textbf{-43.10}&26.72 & \textbf{-27.24}&20.67 \\
 & No. conversions & 0.37&34.91 & -10.29&32.49 \\
 & Cumulative idle time & \textbf{-63.54}&47.56 & \textbf{-53.82}&52.62 \\
 & Completion rate & \textbf{-36.35}&13.73 & \textbf{-26.89}&12.41 \\
\hline
\multirow{4}{*}{High}& Tardiness & \textbf{-12.20}&9.05 & \textbf{-8.67}&8.37 \\
 & No. conversions & \textbf{-92.67}&105.09 & \textbf{-111.33}&83.35 \\
 & Cumulative idle time & \textbf{-109.30}&50.27 & \textbf{-116.08}&56.43 \\
 & Completion rate & \textbf{-35.10}&12.42 & \textbf{-27.64}&11.71 \\
\hline\hline
\end{tabular}
\end{table}

\begin{table}[h]
\caption{Percentage improvement of comparison models in four metrics compared to the proposed model for the short-term production scenario}
\label{table_4}
\centering
\begin{tabular}{c|c|r|r|r|r}
\hline\hline
 \multirow{2}{*}{Demand} & \multirow{2}{*}{Metric} & \multicolumn{2}{c|}{DRL-JSSP} & \multicolumn{2}{c}{DRL-DFJSS}\\
 \cline{3-6}
 &  & Mean & Std. & Mean & Std.\\
 \hline
\multirow{4}{*}{Low}& Tardiness & 9.70&64.39 & -15.32&70.83 \\
 & No. conversions & \textbf{-87.17}&73.15 & \textbf{-78.72}&67.86 \\
 & Cumulative idle time & \textbf{-4.68}&11.39 & \textbf{-3.51}&12.35 \\
 & Completion rate & 0.24&5.04 & -0.12&5.60 \\
\hline
\multirow{4}{*}{Medium}& Tardiness & \textbf{-17.02}&52.34 & \textbf{-6.19}&43.29 \\
 & No. conversions & \textbf{-78.32}&80.20 & \textbf{-74.82}&73.51 \\
 & Cumulative idle time & \textbf{-11.37}&16.72 & \textbf{-12.82}&16.70 \\
 & Completion rate & \textbf{-4.81}&10.99 & \textbf{-1.80}&9.35 \\
\hline
\multirow{4}{*}{High}& Tardiness & \textbf{-15.40}&29.55 & \textbf{-12.36}&29.73 \\
 & No. conversions & \textbf{-59.01}&58.70 & \textbf{-47.09}&50.48 \\
 & Cumulative idle time & \textbf{-21.10}&16.99 & \textbf{-18.37}&18.83 \\
 & Completion rate & \textbf{-8.72}&10.41 & \textbf{-5.84}&11.03 \\
\hline\hline
\end{tabular}
\end{table}

\begin{figure}[h]\centering
\subfloat[Low-demand]{\includegraphics[width=0.32\linewidth]{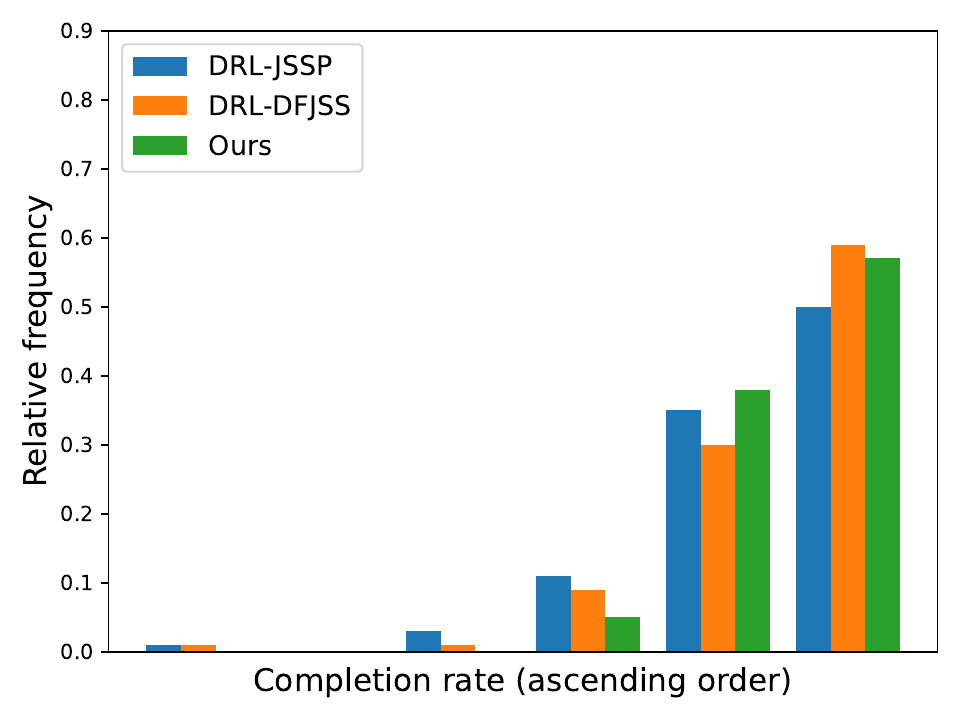}}
\subfloat[Medium-demand]{\includegraphics[width=0.32\linewidth]{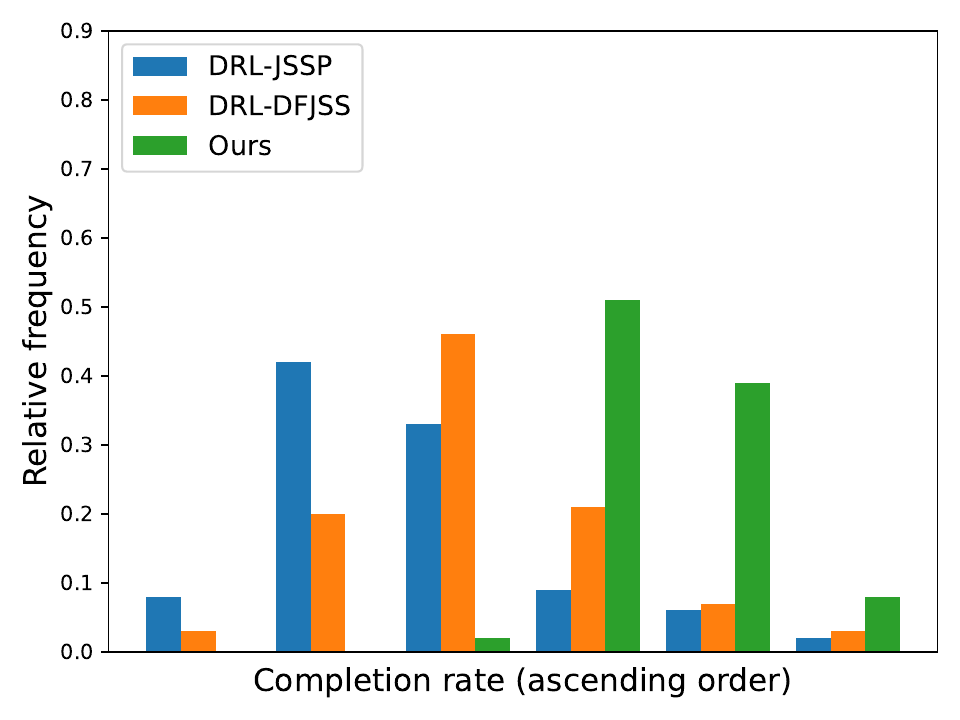}}
\subfloat[High-demand]{\includegraphics[width=0.32\linewidth]{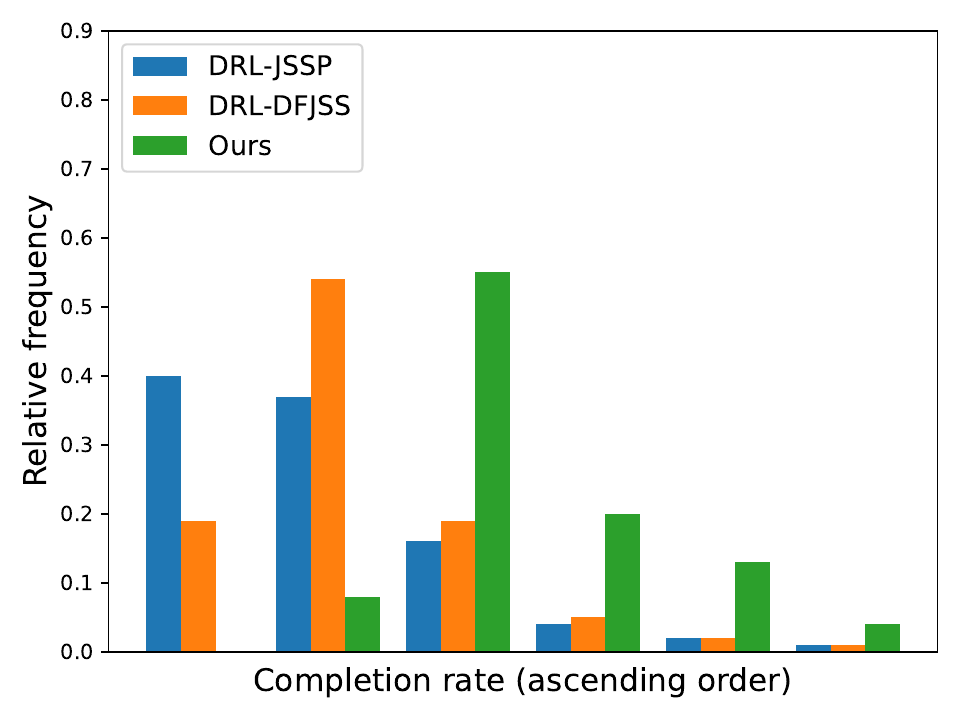}}
\caption{Completion rate histogram of three comparison models for the long-term production scenario. The x-axis tick values are omitted for confidentiality.}
  \label{fig:long_CR}
\end{figure}

\begin{figure}[h]\centering
\subfloat[Low-demand]{\includegraphics[width=0.32\linewidth]{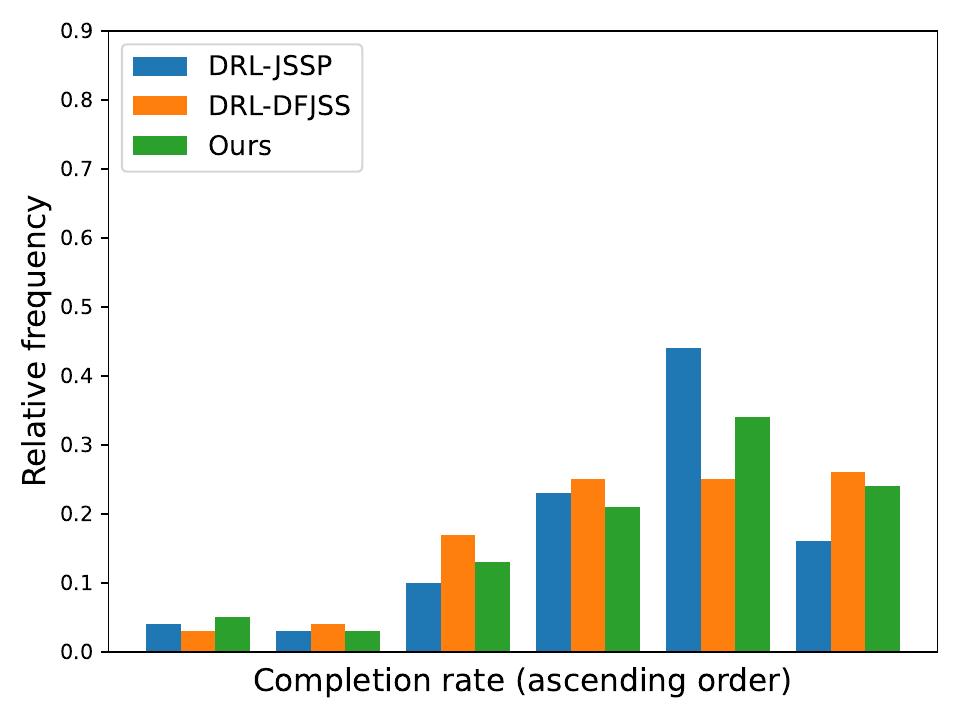}}
\subfloat[Medium-demand]{\includegraphics[width=0.32\linewidth]{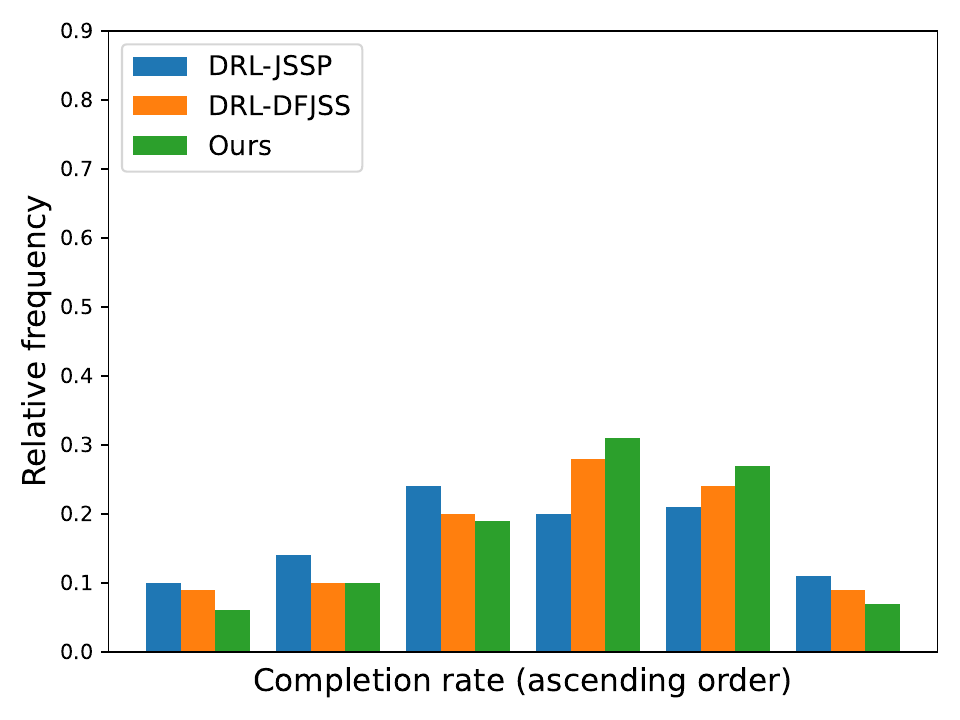}}
\subfloat[High-demand]{\includegraphics[width=0.32\linewidth]{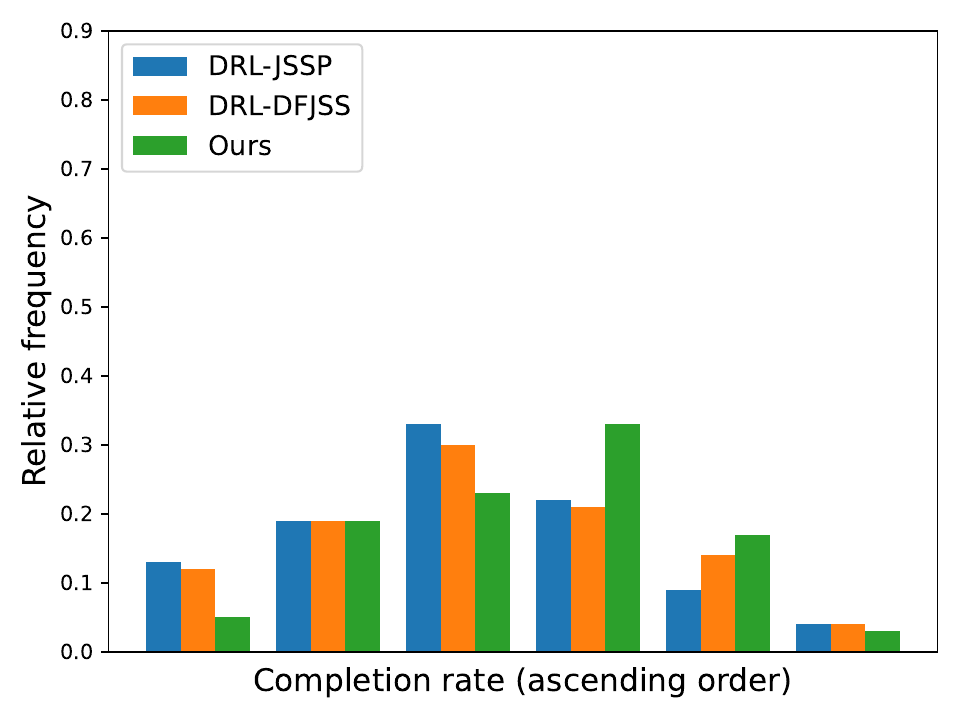}}
\caption{Completion rate histogram of three comparison models for the short-term production scenario. The x-axis tick values are omitted for confidentiality.}
  \label{fig:short_CR}
\end{figure}

We additionally compare the three models based on their completion rates since this metric is closely related to the number of delayed lots used in reward calculations. The results are presented in Figs. \ref{fig:long_CR} and \ref{fig:short_CR}. The proposed model achieves a higher completion rate in most experimental settings, especially demonstrating greater proportions in the higher ranges of the histograms. This indicates a lower risk of poor performance with our model, a trend that becomes more pronounced as demand increases.

\subsection{Ablation study}
To verify the effectiveness of each component proposed in this study, we conduct a quantitative ablation study. Specifically, we generate 100 production plans based on the long-term scenario outlined in Table \ref{table_1}. We then compare the performance of the following models:
\begin{enumerate}[1)]
\item Shared reward model (SRM): Under this model, the followers are updated using a reward-sharing strategy, where the rewards are calculated based on all delayed lots and distributed equally among all followers.
\item Operation-wise reward model (ORM): This model calculates rewards for each follower based on delayed lots of all processable products in the assigned operation.
\item Leader-follower shared reward model (LFSRM): The leader-follower concept using the abstract goal is incorporated into SRM.
\item Leader-follower operation-wise reward model (LFORM): The leader-follower concept is incorporated into ORM. Specifically, LFORM introduces operation-wise reward calculation to the LFSRM framework. 
\item Proposed model: The proposed model, LFORM-RC, incorporates rule-based conversion in addition to the LFORM approach. The acronym RC signifies the inclusion of rule-based conversion as a novel feature of the model.
\end{enumerate}

We establish SRM as the baseline and assess the improvement of the other models based on the four metrics used in Section \ref{comp:sota} compared to the baseline. The comparison results are presented in Table \ref{table_2}. It is important to note that, while a small number of conversions is desirable as it can reduce operating costs, other metrics such as tardiness and completion rate take priority, as they are crucial for achieving high productivity. The results demonstrate that the operation-wise rewarding strategy, the leader-follower concept using abstract goals, and the rule-based conversion contribute to overall performance improvement. In particular, the rule-based conversion is crucial in improving tardiness and completion rates by preventing significant capacity losses through the minimization of poor conversions.

\begin{table}[h]
\scriptsize
\caption{Percentage improvement in four metrics compared to SRM after conducting 100 simulations and averaging the results}
\label{table_2}
\centering
\begin{tabular}{c|r|r|r|r|r|r|r|r}
\hline\hline
 & \multicolumn{2}{c|}{\multirow{2}{*}{ORM}} & \multicolumn{2}{c|}{\multirow{2}{*}{LFSRM}} & \multicolumn{2}{c|}{\multirow{2}{*}{LFORM}} & \multicolumn{2}{c}{LFORM-RC} \\
& \multicolumn{1}{c}{} & \multicolumn{1}{c|}{} & \multicolumn{1}{c}{} & \multicolumn{1}{c|}{} & \multicolumn{1}{c}{} & \multicolumn{1}{c|}{} & \multicolumn{2}{c}{(Ours)}  \\
\cline{2-9} 
 &  Mean & Std. & Mean & Std. & Mean & Std. & Mean & Std.\\
\hline 
Tardiness & 25.20&50.86 & -19.81&117.53 & -6.85&63.98 & \textbf{84.08}&15.42\\
No. conversions & -0.48&37.03 & 11.49&24.49 & 11.22&24.87 & -265.56&106.54\\
Cumulative idle time & 4.27&10.32 & 6.47&13.26 & 3.01&13.16 & \textbf{11.08}&8.92\\
Completion rate & 14.24&23.48 & 8.11&22.46 & 7.45&22.99 & \textbf{33.23}&23.80\\
\hline\hline
\end{tabular}
\end{table}

\section{Conclusion}\label{sec:conclusion}
In this study, we present a novel scalable MARL model for dynamic scheduling problems in real-world factories. The proposed model addresses several hurdles faced in factory scheduling by combining operation-specific agent modeling, abstract goal-based leader-follower coordination, and a rule-based conversion algorithm, which are novel in the field of RL-based scheduling. In particular, we found that RL agents can make errors that lead to significant losses in production capacity. To mitigate this issue, a rule-based conversion algorithm can be used to supplement the agents' operational decisions. Our experiments, conducted on simulators constructed using Intel production datasets, demonstrate that the proposed model significantly outperforms state-of-the-art RL-based scheduling models that rely on human-made dispatching rules. This highlights the limitations of existing methods that use a combination of human-made rules and the potential of our approach to provide more efficient scheduling solutions for real-world factories.

While our proposed model exhibits good performance and scalability, there is an issue that needs to be addressed in our future research. The model assumes that production-related information, such as operation list, machine list, and product types, remain fixed. However, in real-world factories, these settings can change from time to time, necessitating a retraining of the model. Although this can be easily addressed with sufficient computing resources, the process requires frequent manual interventions, which can be a challenge. To overcome this, we need to develop a dynamically evolving model that can adapt to factory setting changes. This direction will be a focus of our future research efforts.

\bibliographystyle{IEEEtran}
\bibliography{reference}\ 

\end{document}